# High performance FeSe$_{0.5}$Te$_{0.5}$ thin films grown at low temperature by pulsed laser deposition

Pusheng Yuan[1,*], Zhongtang Xu[1,*], Haitao Zhang[1], Dongliang Wang[1], Yanwei Ma[1], Ming Zhang[2] and Jianqi Li[2]

[1]*Key laboratory of applied superconductivity, Institute of Electrical Engineering, Chinese Academy of Sciences, Beijing 100190, China*
[2]*Beijing National Laboratory for Condensed Matter Physics, Institute of Physics, Chinese Academy of Sciences, Beijing 100190, China.*

* Yuan P S and Xu Z T contributed equally to this work.

E-mail: ywma@mail.iee.ac.cn

**Abstract**

We report on fully epitaxial FeSe$_{0.5}$Te$_{0.5}$ (FST) thin films with high-quality grown on CaF$_2$ (00*l*) substrate at a low temperature of 300 °C by pulsed laser deposition. The transport $J_c$ of thin films is up to 1.36 MA/cm$^2$ in self-field and 0.97 MA/cm$^2$ in 9 T at 4.2 K, indicating very weak field dependence. A nearly isotropy of $J_c$ ($\gamma = J_c^{H//ab}/J_c^{H//c}$) as low as 1.09 at 9 T is achieved in the FST thin films. Moreover, no clear amorphous interfacial layer presents between the film and the substrate probably ascribing to low temperature and low laser repetition rate grown, while the thickness of reaction layer is approximate 5 nm in many other works. The evidences of transmission electron microscopy show some lattices with lateral size of < 5 nm×20 nm seem to be disturbed. Those location defects are thought to be responsible for nearly isotropic behavior of superconductivity.

## 1. Introduction

After the discovery of iron-oxypnictide superconducting materials [1], Hsu et al reported superconductivity in the FeSe with a transition temperature of 8 K [2], which is now called "11" family. Superconductors of "11" occupy an important position in the emerging iron-based superconductors because of their simplest structures, no arsenic involved, and relatively easy synthesis. Furthermore, their superconducting transition temperature $T_c$ can be greatly enhanced by external pressure or the substrate-induced strain in thin films. The $T_c$ of $FeSe_{0.5}Te_{0.5}$ (FST) films can be increased to 21 K due to the strain caused by the substrates which is much higher than that of bulk material [3]. Therefore, superconductors of "11" family have generated a tremendous interest in the scientific community. FST films have been successfully grown on various single-crystalline substrates or coated conductor substrates in last five years. However, an oxygen-rich interfacial layer between FST films and oxide substrates is always formed [4], which deteriorates the superconducting properties. Transmission electron microscope (TEM) analysis indicated oxygen diffusion from oxide substrates into the films and formed interfacial layer [5]. Alternatively, choice of non-oxidizing substrates or oxide substrates with a proper buffer layer is proved to be an effective way to prevent the oxygen diffusion and thus improve the performances of FST thin films. Although utilizing suitable buffer layer can effectively avoid the reaction layer and increase the $J_c$ of FST films [6, 7], the intrinsic physical properties of the films may be influenced or shielded. Hence, non-oxidizing substrates may be more suitable for Fe-based superconducting thin films which have been already supported by films grown on $CaF_2$ substrate with high crystalline and excellent superconducting properties [8-11]. To date, at least four groups have reported FST films successfully prepared on $CaF_2$ substrates by pulsed laser deposition (PLD) [9, 12-14]. The fabrication temperature of the substrate varied from 280 to 600 ℃. The FST films grown at relatively low temperature (< 300 ℃) usually show a low $J_c$ while the films with the highest $J_c$ (4 K, 0T) > 1.1 MA/cm$^2$ and the highest $T_c^0$=19 K are grown on $CaF_2$ substrates at a high temperature of 550 ℃ [12]. Furthermore, FST films grown on $CaF_2$ substrates at both high and low-temperature, amorphous interfacial layer often appears in the interface of film and substrate. The interfacial layer not only deteriorates the superconducting properties [5] but also disturbs effective device control for electronic sensor applications [15].

In the present work, we report FST thin films with high $J_c$ and high $T_c$ grown on $CaF_2$ substrates without apparent amorphous interfacial layer at a relatively low temperature of 300 ℃. We show FST films exhibit $T_c^{zero}$ = 18.8 K and $J_c$ values at 4.2 K of 1.36 MA/cm$^2$ in self-field and 0.97 MA/cm$^2$ at 9 T. These $J_c$ values are the highest ever reported so far in the 11 phase films.

**2. Experimental details**

FST thin films were grown by PLD using a KrF excimer laser (wavelength: 248 nm). The target $FeTe_{0.5}Se_{0.5}$ with $T_c^{zero}$ = 14 K was prepared by solid state reaction as reported in reference [13]. The deposition temperature was 300 ℃ and the pressure was kept lower than 10$^{-7}$ Torr; the laser energy was set at 320 mJ/pulse on the target and the laser repetition rate was 3 Hz; the distance between substrate and target was set to be 45 mm. Structural characterizations and phase purity of the films were characterized by X-ray diffraction (XRD) with Cu Kα radiation at room temperature. Taking into account of the field was applied perpendicular and parallel to the $c$-axis, the relationship between $J_c$ of the film and magnetic field was investigated at 4.2 K in a field range of 0–9 T by Physical Properties Measuring System (PPMS, Quantum Design). In order to reduce temperature drift during the testing process caused by the heat, micro bridges of 20 μm in width and 100 μm in length were fabricated by conventional photolithography and Ar$^+$ etching. To characterize the microstructure of the films, TEM and high-resolution TEM (HRTEM) were used. The thickness of FST films are about 200 nm determined by scanning electron microscope (SEM).

**3. Results and discussion**

Figure 1(a) shows XRD patterns for thin films normalized by the intensity of the FST (00$l$) peaks. Besides from the diffraction peaks of the substrate, only sharp and strong (00$l$) peak diffractions assigned to FST thin films are observed, indicating $c$-axis texture and high phase purity. The full width at half maximum (FWHM) of the (002) peak calculated from the rocking curve shown in the inset of Fig. 1 is 1.275°, slightly higher than previously reported FST films on $CaF_2$ substrates [16] and other substrates [3, 7]. This may be caused by the low deposition temperature. The value of c-axis lattice constant extracted from fitting the (004) peaks is 5.93 Å, much smaller than that of the bulk (about 6.01 Å), but are comparable to the FST films in many other works. Figure 1 (b) shows the φ scan of (101) peak from the FST film and (202) peak from the substrate. It exhibits no satellite or additional diffractions besides four peaks of film and substrate with 90° intervals, indicative of a fourfold symmetry and biaxial texture. Peaks of the films are separated from that of the substrate by a rotation of 45°, indicating the epitaxial growth of the films on the substrates with a relationship of (001)[100] film//(001)[110] substrate. It is the same as Co-doped Ba-122 thin films grown on versatile fluoride substrates [17]. The FWHM of the φ scan peaks is approximately 0.39°, only half of the sample grown at 550 ℃ [12] and one-third smaller than the sample grown at 300 ℃ [13]. The preceding XRD results indicate that FST films have good crystalline quality and high phase purity.

The temperature dependence of the resistivity was measured, as shown in Fig.2, by the physical PPMS with a standard four-probe method. The FST film shows metallic over the whole temperature range and exhibits a very sharp resistive transition. The onset superconducting transition temperature and the zero resistance temperature are $T_c^{onset}$=19.5K and $T_c^{zero}$=18.8 K, respectively, with a narrow transition width about 0.7 K as shown in the inset, suggesting high quality of the films. The $T_c$ enhancement behavior in FST films is reminiscent of the giant pressure effects in iron-based superconductors. Using the relationship $1/d_{101}^2 = 1/a^2 + 1/c^2$, we calculated the a-axis value of the film is about 3.78 Å. This value is smaller than that of the bulk (3.81 Å) and the $CaF_2$ substrate (3.86 Å). The unit cell volume of the film is nearly 3% smaller than the bulk. The lattice deformation may be caused by the strain effect due to the mismatch between the films and the substrates [3] or chemical substitution [18]. As a magnetic field of 9 T is applied to the film, a clear shift of $T_c$ to lower temperatures from 18.8 K (in self-field) to 16.6 K (H//*c*) and 17.3 K (H//*ab*) is observed. Figure 2(b) shows the upper critical fields $H_{c2}$ and irreversibility fields $H_{irr}$ of FST films, which are estimated with the criteria of 90% and 10% of resistivity at normal state, respectively. In addition, using Werthamer–Helfand–Hohenberg (WHH) formula: $H_{c2}(0)$ =-0.693$T_c$(d$H_{c2}$/dT), the $H_{c2}$ at zero-temperature can be deduced. The values of $H_{c2}(0)$ are 121 and 76 T for (H//*c*) and (H//*ab*), respectively. In brief, the superconductivity of the films grown on $CaF_2$ substrates at a relatively low temperature (300 °C) is robust against the magnetic field.

According to thermally activated flux flow theory, the relation between the resistivity and pinning potential of iron-based superconductor can be written as Arrhenius equation [19]:

$$\rho = \rho_0(H) e^{-\frac{U_0(H)}{K_B T}}$$

Arrhenius plots of resistivity as $\ln\rho$ vs. $T^{-1}$ of the resistivity data for both directions (H// $ab$ and H// $c$) measured in static fields up to 9 T shows a quite wide linear region, indicating the sustainability of the above cited approximation. The pinning potential $U_0(H)/K_B$ was evaluated from the linear fit as a function of the applied magnetic field, as shown in Fig. 3 (a). The $U_0(H)/K_B$ value of H parallel to $ab$ plane is higher than that of H parallel to $c$ axis, but they have a similar tendency in magnetic field. A single exponent is not sufficient to fit the data by a power law $U_0(H)/K_B \propto H^{-\alpha}$ in the whole magnetic field range. With magnetic field increases to 1.5 T, the data is well fitted by $\alpha = 0.12$ and 0.15 for the field parallel the $ab$ planes and parallel the $c$ planes, respectively. Whereas $\alpha = 0.36$ and 0.4 are obtained when the magnetic field increases from 1.5 to 9 T, which are comparable to previously reported work [20]. In the whole range of applied magnetic field (0-9 T), the $\alpha$ value is less than 0.5 which means weak dependence on the magnetic field. The best fit of the experimental data implies the value of the pinning potential ($U_0/K_B$) in 0.5 T is lower than that of $MgB_2$ [21]. As the magnetic field increases to higher than 3 T, the pinning potential is greater and decreases more slowly than $MgB_2$. This result indicates that the FST films may be a competitive candidate to replace low Tc superconductors for high-field application at liquid helium temperature.

The transport measurements for FST film as a function of the magnetic field in two directions (H//$ab$ and H//$c$) at 4.2 K is shown in Fig. 3(b). $J_c$ is $1.36 \times 10^6 A/cm^2$ in self-filed, which is higher than that of the films grown on $CaF_2$ substrates at 550 °C [12] and 300 °C [13]. Meanwhile, $J_c$ is still maintained $0.97 \times 10^6 A/cm^2$ (H//$ab$) and $0.89 \times 10^6 A/cm^2$ (H//$c$) at 9 T, indicating a small anisotropy of $J_c$($\gamma = J_c^{H//ab}/J_c^{H//c} = 1.09$) and very weak field dependence. The pinning force is 87.3 $GN/m^3$ (H//$ab$) and 80.1 $GN/m^3$ (H//$c$), which are the highest value for 11 family [6, 7, 12, 13].

The TEM studies were performed on the thin films to investigate the correlation between microstructure and pinning mechanisms. TEM characterization proves that the FST films have high quality and are absent of apparent reaction layer. Figure 4(a) illustrates bright-field TEM image of cross-sectional FST films. Atomic arrangement orientation of film rotated by 45 degrees relative to the substrate can be found, which is consistent with the φ scan result. At the same time, there is a sharp interface between FST film and the substrate without amorphous layer or any remarkable lattice disorder. In contrast, FST films synthesized at high temperature of 550 °C present approximately 5 nm thickness of reaction layer [12]. On the other hand,, there is a rather rough interface between the $CaF_2$ substrate and the FST films grown at the temperature of 280 °C [18, 22].Those growth temperature is 20 °C lower than our investigation, but the laser repetition rate is 10 Hz that is more than three times larger than that of our work. The laser repetition rate influence the growth rate of the film, and consequently affects the microstructure of thin films and interfaces [23]. It is important that low temperature (300 °C) and low laser repetition rate fabrication process inhibit the interfacial layer of FST films grown on $CaF_2$ substrates without buffer layer. Some local lattice seems to be disturbed with lateral size of less than 5 nm×20 nm as shown in Fig. 4(b). Such defects might originate from the departure of Se and Te atoms from the local stoichiometry, because the occupancy of Se and Te atoms at different positions in the unit cell will give rise to a splitting of Se/Te in the c-axis direction of the films and then lead to the atomic scale disorder, as confirmed by STM imaging [24]. Those disturbed and randomly orientated lattices are different from large-extended defects throughout the whole film as found in FST films grown on $LaAlO_3$ substrates and $BaFe_2(As,P)_2$ films grown on MgO substrates [23]. The anisotropy of $J_c$ is less than 1.09 in the whole applied magnetic field region, which may be associated with this type of defect. The inset of Fig. 4(b) is the selected area electron diffraction (SAED) from FST films and $CaF_2$ substrate. The SAED pattern results indicate the region was not amorphous. According to the inset of Fig. 4(b), the lattice constants of the a- and c-axis of the FST film are 3.77 and 5.93 Å, respectively, which are comparable to the results of the XRD.

## 4. Conclusions

A self-field $J_c$ = 1.36 MA/cm$^2$ and $J_c$ over 0.97 MA/cm$^2$ in 9 T were obtained for epitaxial FST films grown on CaF$_2$ substrate without any buffer layer. These are the highest $J_c$ values in FST films. More importantly, FST films present no apparent reaction layer because of low fabrication temperature. Local defects without clear orientation were observed in FST films, which may be responsible for small anisotropy of $J_c$ in the whole applied magnetic field region. Our work proves that FST films directly grown on the substrate and at low temperature are high quality, absence of reaction layer and high current transport performance.

**Acknowledgements**

The authors would like to express their thanks to Prof. Dongning Zheng and Dr. Hui Deng for help in the fabrication of micro bridges. This work was supported by the National "973" Program (Grant No. 2011CBA00105), National Natural Science Foundation of China (Grant Nos.51172230, 51320105015) and the Beijing Municipal Science and Technology Commission (No.Z141100004214002).

**Captions**

Fig. 1 (a) XRD pattern of FeSe$_{0.5}$Te$_{0.5}$ thin film deposited on CaF$_2$ substrate. The inset shows ω-scan rocking curve of the (002) reflection. (b) The φ scan of (101) peak from FeSe$_{0.5}$Te$_{0.5}$ and (202) peak from CaF$_2$ substrate, lattice alignment of the film rotated 45 degrees relative to the substrate.

Fig. 2 (a) The resistance versus temperature of FeSe$_{0.5}$Te$_{0.5}$ films on CaF$_2$ substrate. The inset shows temperature dependent resistance for FeSe$_{0.5}$Te$_{0.5}$ films in different magnetic fields. (b) Upper critical field H$_{c2}$(T) and irreversibility field H$_{irr}$(T) of FeSe$_{0.5}$Te$_{0.5}$ films on CaF$_2$ substrates with the field parallel *ab* plane and *c* axis, respectively.

Fig. 3 (a) Vortex motion activation energy as a function of field in parallel *ab* plane and parallel *c* axis configurations as extracted from the Arrhenius plot. (b) Critical current density as a function of magnetic-field for the film deposited on CaF$_2$ substrate for H //*ab* and H //*c*. Closed and open symbols show the data obtained by applying H// *ab* and H// *c*, respectively

Fig.4 (a) HRTEM images of the interfacial layer between the CaF$_2$ substrate and FeSe$_{0.5}$Te$_{0.5}$ thin film. The dashed red line marks the interface between FeSe$_{0.5}$Te$_{0.5}$ film and the substrate. (b) Some local defects (marked in red ellipse) present in the FST films, and the inset is SAED image of FeSe$_{0.5}$Te$_{0.5}$ thin films on CaF$_2$ substrate.

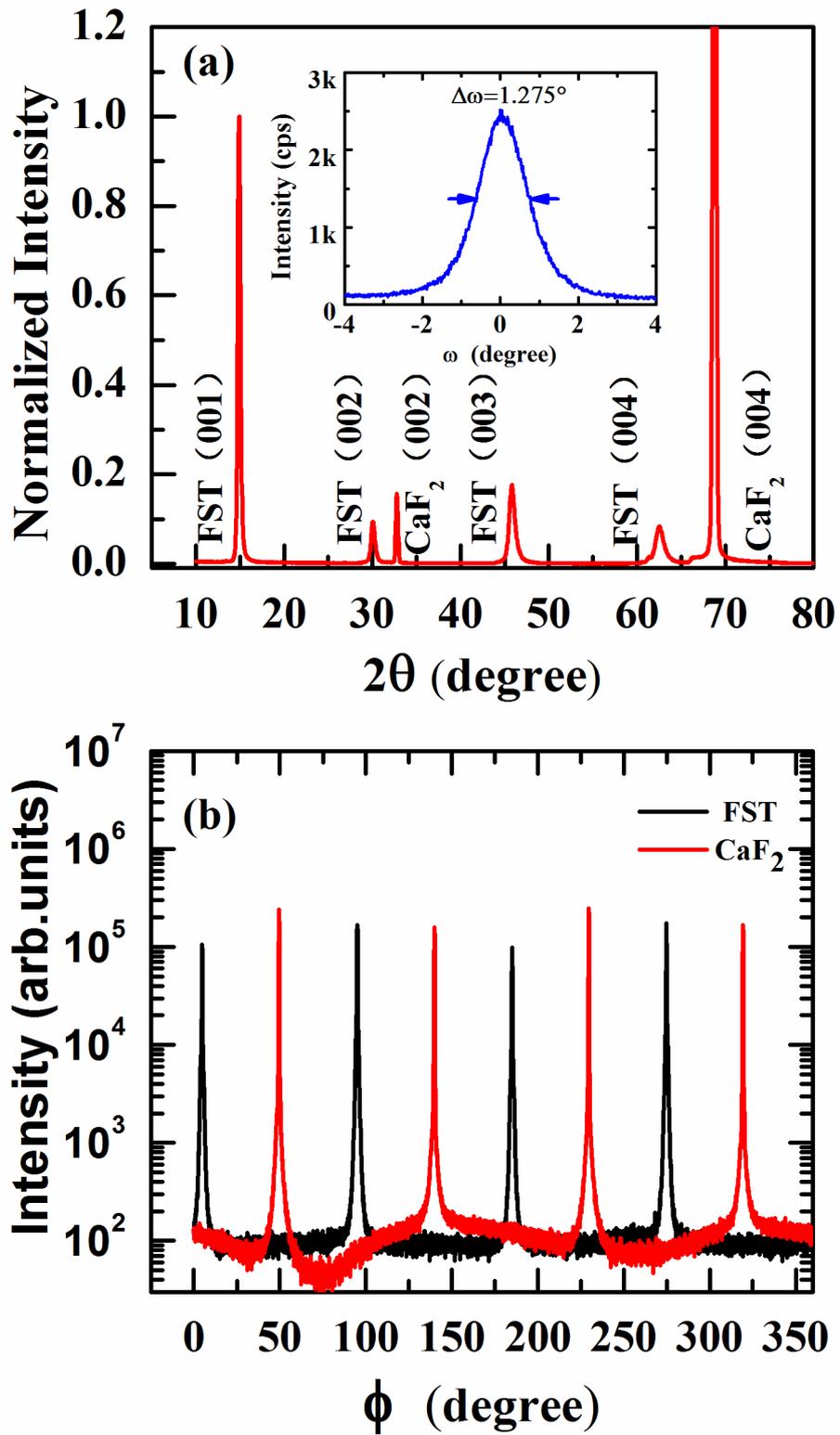

Fig. 1. P.S.Yuan *et al*.

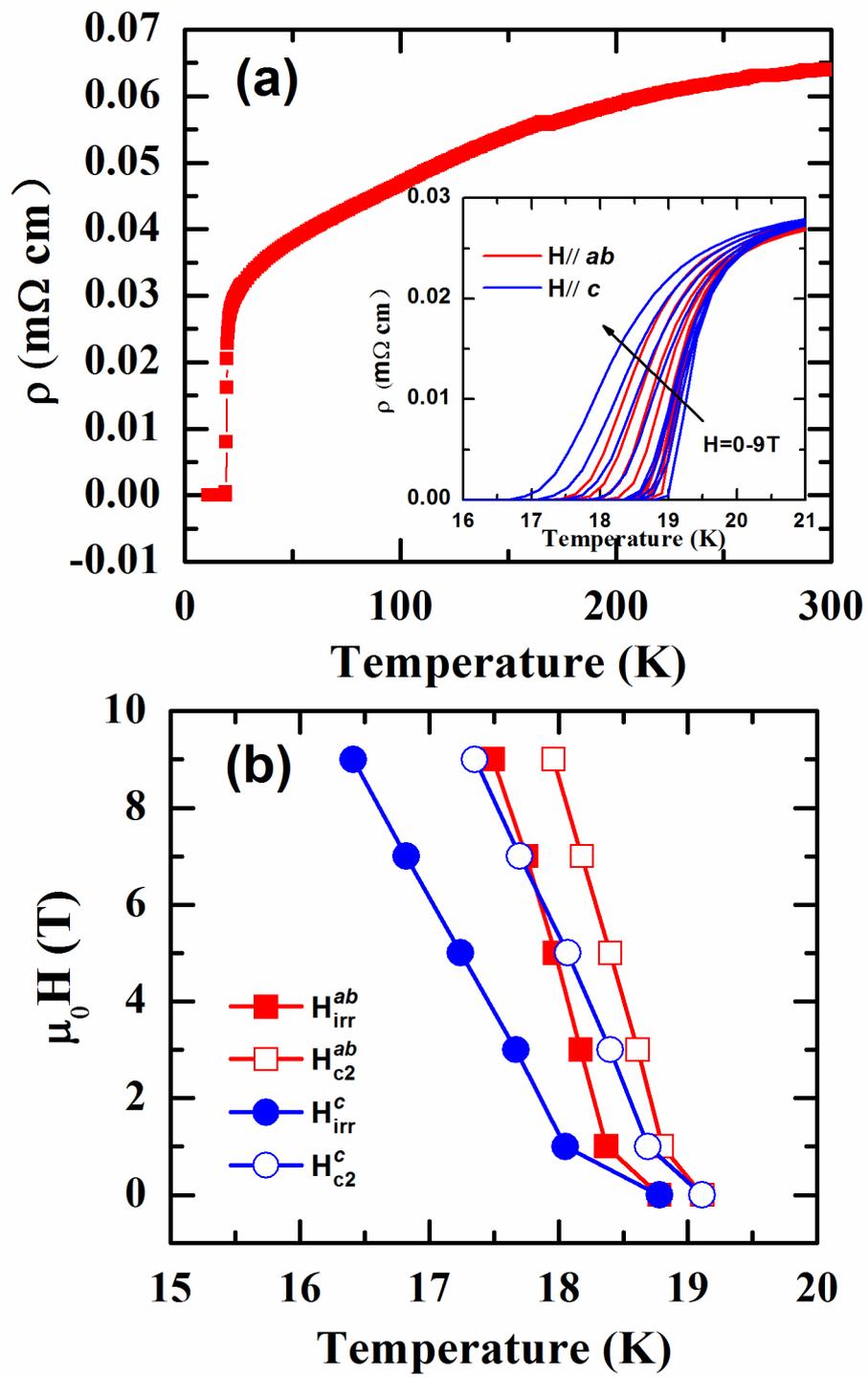

Fig. 2. P.S.Yuan et al.

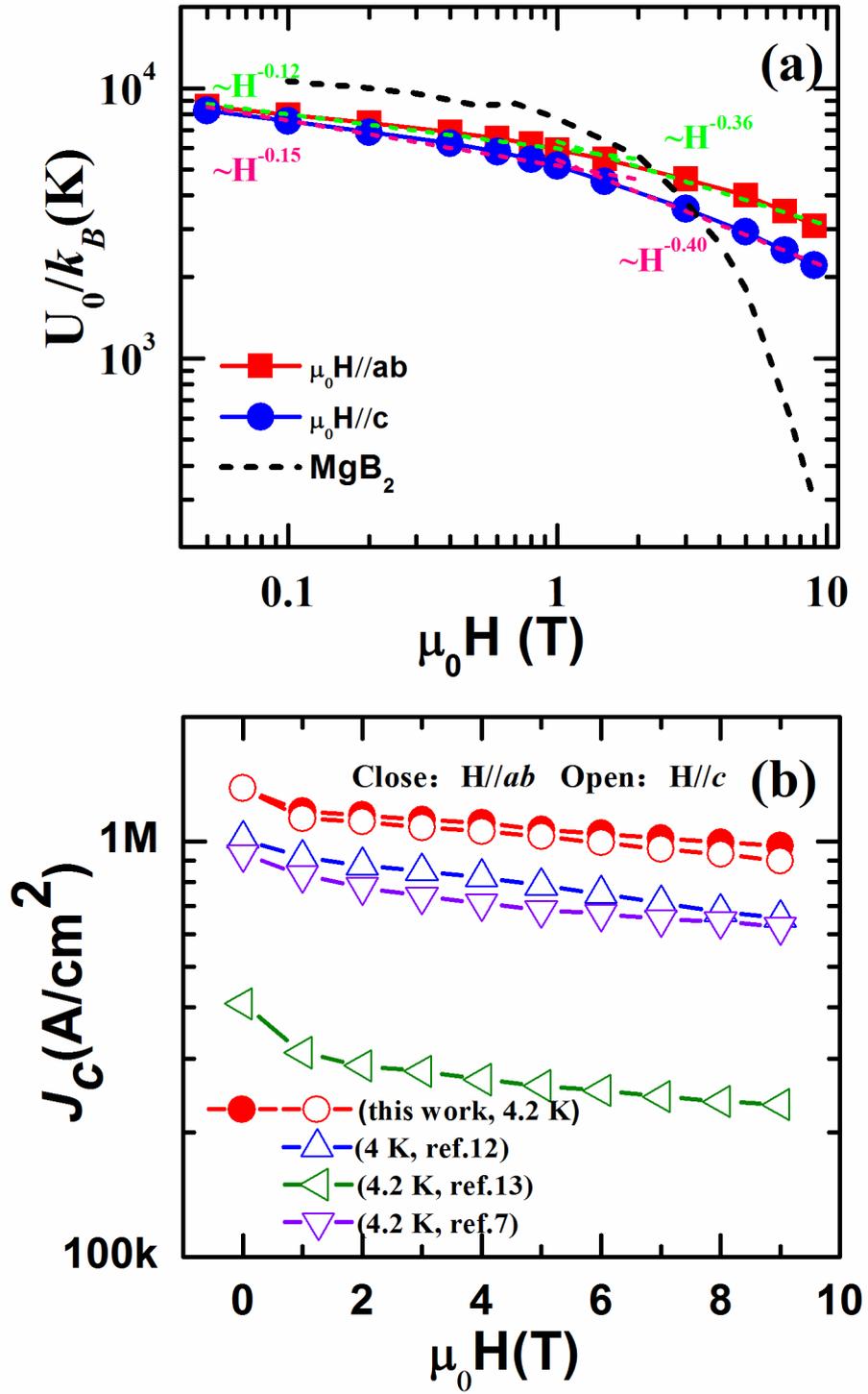

Fig. 3. P.S. Yuan et al.

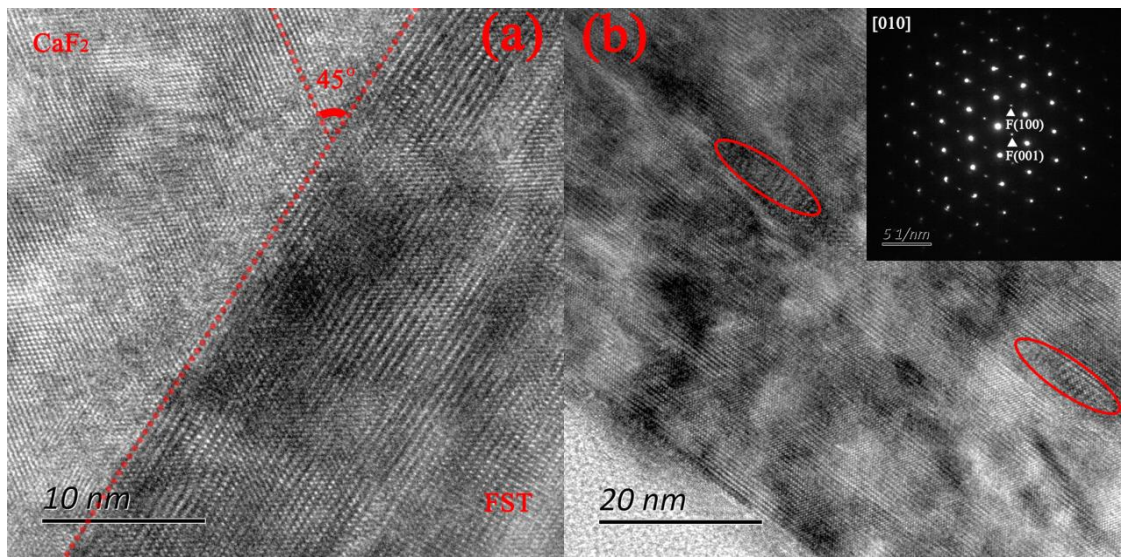

Fig. 4. P.S.Yuan et al.